\documentclass{sig-alternate}
\title{Intention-Oriented Process Model Discovery from Incident Management Event Logs}

\numberofauthors{1}

\author{
\alignauthor{
Ashish Sureka \\
\affaddr{Software Analytics Research Lab (SARL), India \\
www.software-analytics.in} \\
\email{ashish@iiitd.ac.in}
}
}

\begin{document}
\maketitle
\begin{abstract}
Intention-oriented process mining is based on the belief that the fundamental nature of processes is mostly intentional (unlike activity-oriented process) and aims at discovering strategy and intentional process models from event-logs recorded during the process enactment. In this paper, we present an application of intention-oriented process mining for the domain of incident management of an Information Technology Infrastructure Library (ITIL) process. We apply the Map Miner Method (MMM) on a large real-world dataset for discovering hidden and unobservable user behavior, strategies and intentions. We first discover user strategies from the given activity sequence data by applying Hidden Markov Model (HMM) based unsupervised learning technique. We then process the emission and transition matrices of the discovered HMM to generate a coarse-grained Map Process Model. We present the first application or study of the new and emerging field of Intention-oriented process mining on an incident management event-log dataset and discuss its applicability, effectiveness and challenges.  
\end{abstract}
\section{Research Motivation and Aim}
Process mining consists of extracting or discovering run-time process models from event logs, measuring the extent of compliance between the design time and actual process models, improving process models and analyzing it from multiple perspectives (such as control-flow and organizational perspectives) \cite{vanderAalst2004}\cite{vanderAalst2011}. Intention-oriented process mining is a new and emerging area which is based on the belief that the fundamental nature of processes is mostly intentional (unlike activity-oriented process mining which specifies behaviors in-terms of sequences of tasks and branches) \cite{Khodabandelou2014}\cite{Khodlncs2013}\cite{Khodlncs2014}. Intention-oriented process mining is a relatively unexplored area except the recent (year $2013$ and $2014$) and pioneering work on the topic by Khodabandelou et al. \cite{Khodabandelou2014}\cite{Khodlncs2013}\cite{Khodlncs2014}.  Discovery of intentional process models from event logs has been proposed for the first time by Khodabandelou et al. and we believe that several more studies are required to throw more light on the applicability and effectiveness of intention-oriented process modeling for solving practical problems encountered by process owners in enterprises. Khodabandelou et al. demonstrate the application of intention-oriented process modeling on two datasets: event logs of developers who committed their activities to Usage Data Collector (UDC) of Eclipse and a laboratory context where an experiment was conducted with University Master students for Entity-Relationship diagrams creation \cite{Khodabandelou2014}\cite{Khodlncs2013}\cite{Khodlncs2014}. The study presented in this paper is motivated by the need to extend the state-of-the-art in the emerging area of intention-oriented process modeling by applying the approach on a large real-world dataset and sharing the findings with the research community. The specific research aims of the work presented in this paper is the following:
\begin{enumerate}
\item To investigate the application of intention-oriented process modeling approach on real-world event-log data extracted from incident management systems of an enterprise. To the best of our knowledge, this is the \textit{first study} on the application of intention mining and intention-oriented process model discovery on a large real-world incident management dataset.  
\item To examine customization or extention of the approach proposed by the pioneers of intention-oriented process modeling for the given context and application scenario.   
\end{enumerate}
\section{Experimental Dataset}
\begin{figure}[t]
\centering
\includegraphics[width=0.48\textwidth]{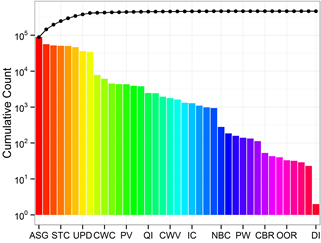}
\vspace{-0.70cm}
\caption{Pareto chart showing the distribution of activities and their cumulative count. Y-axis is in logarithmic scale. }
\label{fig:pareto}
\end{figure}
\begin{table}[t]
\caption{Actor, Activity and Timestamp for one of the Cases in the Dataset}\label{tab:case} 
\begin{center}
\begin{tabular}{|c|l|c|}
\hline
\scriptsize \textbf{DateStamp} & \scriptsize \centering \textbf{Activity} & \scriptsize \textbf{Group} \\ \hline
\scriptsize 7/1/2013 8:17 & \scriptsize Reassignment & \scriptsize 01 \\ \hline
\scriptsize 4/11/2013 13:41 & \scriptsize Reassignment & \scriptsize 02 \\ \hline
\scriptsize 4/11/2013 13:41 & \scriptsize Update from cust & \scriptsize 02 \\ \hline
\scriptsize 4/11/2013 12:09 & \scriptsize Operator Update & \scriptsize 03 \\ \hline
\scriptsize 4/11/2013 12:09 & \scriptsize Assignment & \scriptsize 03 \\ \hline
\scriptsize 4/11/2013 13:41 & \scriptsize Assignment & \scriptsize 02 \\ \hline
\scriptsize 4/11/2013 13:51 & \scriptsize Closed & \scriptsize 03 \\ \hline
\scriptsize 4/11/2013 13:51 & \scriptsize Caused By CI & \scriptsize 03 \\ \hline
\scriptsize 4/11/2013 12:09 & \scriptsize Reassignment & \scriptsize 03 \\ \hline
\scriptsize 25/09/2013 08:27 & \scriptsize Operator Update & \scriptsize 03 \\ \hline
\end{tabular}
\end{center} 
\end{table}
We conduct our study on large real-world publicly available dataset so that our experiments can be replicated and the results can be used for comparison or benchmarking purposes. The work presented in this paper holds the required \textit{replication standards} ensuring sufficient information for any third party to replicate the results without any additional information from us. We conduct experiments on the publicly available dataset provided by the tenth\footnote{ http://www.win.tue.nl/bpi/2014/challenge} International Workshop on Business Process Intelligence (BPI). Data collection is one of the most important stage in conducting qualitative research and the quality of result obtained depends both on research design and data gathered. The data provided on the BPI workshop website is of high quality as it is peer-reviewed and prepared by experts on the given topic. As an academic, we believe and encourage academic code or software sharing in the interest of improving \textit{openness and research reproducibility}. We release our code and dataset in public domain so that other researchers can validate our scientific claims and use our tool for comparison or benchmarking purposes (and also reusability and extension). Our code and is hosted on GitHub\footnote{Currently not mentioned due to blind review policy}[Currently not mentioned due to blind review policy] which is a popular web-based hosting service for software development projects. We select GPL license (restrictive license) so that our code can never be closed-sourced.
\section{Empirical Analysis}
\subsection{Discovering User Strategies from Activities}
The Rabobank Group ICT Incident Dataset consists of $46616$ incidents or cases and $466737$ events. The fields in the event-log dataset are: Incident ID, TimeS-stamp, Incident Activity Number, Incident Activity Type, Assignment Group and KM number (a number related to knowledge document). Table \ref{tab:case} shows the Actor, Activity and Timestamp for one of the Cases in the Dataset. The even-log data in Table \ref{tab:case} shows that several activities are performed by various actors during the workflow and process enactment. Our objective is to apply the Map Miner Method (MMM) that aims at discovering intentions and strategies from event or activity logs and thereby build the actual intention-oriented process model \cite{Khodabandelou2014}\cite{Khodlncs2013}\cite{Khodlncs2014}. The intention-oriented process model is a tool that complements the activity-oriented process model helping the user to better understand the deep nature of the business processes from multiple perspectives. The first step in the process is to model users' strategies in terms of observed activities (present in event logs) using Hidden Markov Models (HMMs). The input to the MMM is the temporal set of user's activities (an activity is an interaction of the user with the information system) during a time-slice wherein the time-slice for a temporal-set or sequence of activity is a specific process trace or case. Figure \ref{fig:pareto} displays a Pareto chart showing the total number of activities and the distribution of $39$ different kind of activities in the dataset (The Y-axis of Figure \ref{fig:pareto} is in logarithmic scale). The activity distribution is skewed as activities such as Assignment (ASG), Operator Update (OU), Reassignment (RASG), Closed (CLD) and Status Change (SC) have $88502$, $56292$, $51961$, $50145$ and $50914$ entries respectively whereas activities such as Referred (REF), Problem Closure (PC), OO Response (OOR), Dial-In (DI) and Contact Change (CC) have  $29$, $40$, $33$, $2$ and $32$ entries respectively. Hidden Markov Models (HMMs) are stochastic finite automaton and statistical Markov Models in which the system being modeled is assumed to be a Markov process with unobserved (hidden) states and can be presented as the simplest Dynamic Bayesian Network\footnote{http://en.wikipedia.org/wiki/Hidden\_Markov\_model}. We use Jahmm\footnote{http://code.google.com/p/jahmm/} library which is a Java implementation of Hidden Markov Model (HMM) related algorithms. Jahmm provides implementation of the Viterbi, Forward-Backward, Baum-Welch and K-Means algorithms in addition to other algorithms related to HMM. We transform the dataset in the format required by the Jahmm library. The source code for data transformation and analysis is available on the GitHub website for the project\footnote{Currently not mentioned due to blind-review policy}[Currently not mentioned due to blind-review policy].  

\begin{table*}[t]
\caption{Table showing the mapping between the underlying $12$ user strategies and activities derived from user's traces recorded during the process enactment }\label{tab:strategy} 
\begin{center}
\begin{tabular}{|c|c|p{11.5cm}|l|}
\hline
\scriptsize $\mathbb{S}$& \scriptsize $\pi$ & \scriptsize \centering \textbf{Activities} & \scriptsize \textbf{Distribution} \\ \hline
\scriptsize $\mathbb{S}_1$ & \scriptsize $0.07$ & \scriptsize Reassignment, Communication with vendor & \scriptsize $[0.97, 0.03]$ \\ \hline
\scriptsize $\mathbb{S}_2$ & \scriptsize $0.19$ & \scriptsize Update from customer, Notify By Change, Open & \scriptsize $[0.08, 0.01, 0.92]$ \\ \hline
\scriptsize $\mathbb{S}_3$ & \scriptsize $0.09$ & \scriptsize External Vendor Assignment, Operator Update, Urgency Change, Communication with customer & \scriptsize $[0.06, 0.83, 0.02,0.09]$ \\ \hline
\scriptsize $\mathbb{S}_4$ & \scriptsize $0.16$ & \scriptsize Assignment & \scriptsize $[1]$ \\ \hline
\scriptsize $\mathbb{S}_5$ & \scriptsize $0.17$ & \scriptsize Closed, Resolved, Quality Indicator Set & \scriptsize $[0.93, 0.03, 0.04]$ \\ \hline
\scriptsize $\mathbb{S}_6$ & \scriptsize $0.11$ & \scriptsize Caused By CI, Reopen & \scriptsize $[0.93, 0.07]$ \\ \hline
\scriptsize $\mathbb{S}_7$ & \scriptsize $0.07$ & \scriptsize Impact Change, Quality Indicator Fixed, Update & \scriptsize $[0.03, 0.17, 0.80]$ \\ \hline
\scriptsize $\mathbb{S}_8$ & \scriptsize $0.10$ & \scriptsize Status Change, Mail to Customer & \scriptsize $[0.93, 0.07]$ \\ \hline
\scriptsize $\mathbb{S}_9$ & \scriptsize $0.01$ & \scriptsize External update, Pending vendor, Problem Closure, Callback Request & \scriptsize $[0.20, 0.78, 0.01, 0.01]$ \\ \hline
\scriptsize $\mathbb{S}_{10}$ & - & \scriptsize Analysis-Research, Description Update & \scriptsize $[0.18, 0.82]$ \\ \hline
\scriptsize $\mathbb{S}_{11}$ & - & \scriptsize Vendor Reference Change, Quality Indicator, Vendor Reference, Incident reproduction & \scriptsize $[0.04, 0.69, 0.26, 0.01]$ \\ \hline
\scriptsize $\mathbb{S}_{12}$ & - & \scriptsize alert stage 1, Service Change & \scriptsize $[0.42, 0.58]$ \\ \hline
\end{tabular}
\end{center} 
\end{table*}
 
We first learn an initial HMM using K-Means learning algorithm. The K-Means learner is initialized with $12$ states (number of strategies is equal to the number of states). We specified the number of states as $12$ (a heuristic, number of states equal to $\frac{1}{3}$ of the number of different kind of activities) as BWA requires the cardinality of the strategy set $\mathbb{S}$. We then pass the initial HMM to the Baum Welch Learner which is the most commonly used algorithm to learn the parameters of a HMM  (unsupervised learning). The model parameters to be learnt are $\mathbb{H=\{E,T\}}$. $\mathbb{T}$ is a list of lists or a square $N$X$N$ matrix, whose $(i,j)$ entry gives the probability of transitioning from state $i$ to state $j$. $\mathbb{E}$ a list of $N$ lists or a matrix with $N$ rows, such that $\mathbb{E}$$[i,k]$ gives the probability of emitting symbol $k$ while in state $i$. The behavior of the HMM is captured in $\mathbb{H}$. We consider a uniform distribution for the probabilities of starting in each initial state. A set of activities that is realized to fulfill a given intention is called as a strategy and thus a strategy comprises of several activities. 

Table \ref{tab:strategy} reveals the discovered mapping between the underlying $12$ user strategies and activities derived from user's traces recorded during the process enactment. As shown in the Table \ref{tab:strategy}, activities Update from customer, Notify By Change and Open constitutes the strategy $\mathbb{S}_2$. Similarly, activity Assignment constitutes the strategy $\mathbb{S}_4$. Table \ref{tab:strategy} shows the emission probabilities for activities within a strategy. The distribution column shows how many times a given activity appears in a given strategy. The related probabilities are called emission probabilities. For example, The hidden state $\mathbb{S}_6$ generates the activities Caused by CI and Reopen with a probability distribution of $0.93$ and $0.07$ respectively. 
\begin{figure}[t]
\centering
\includegraphics[width=0.48\textwidth]{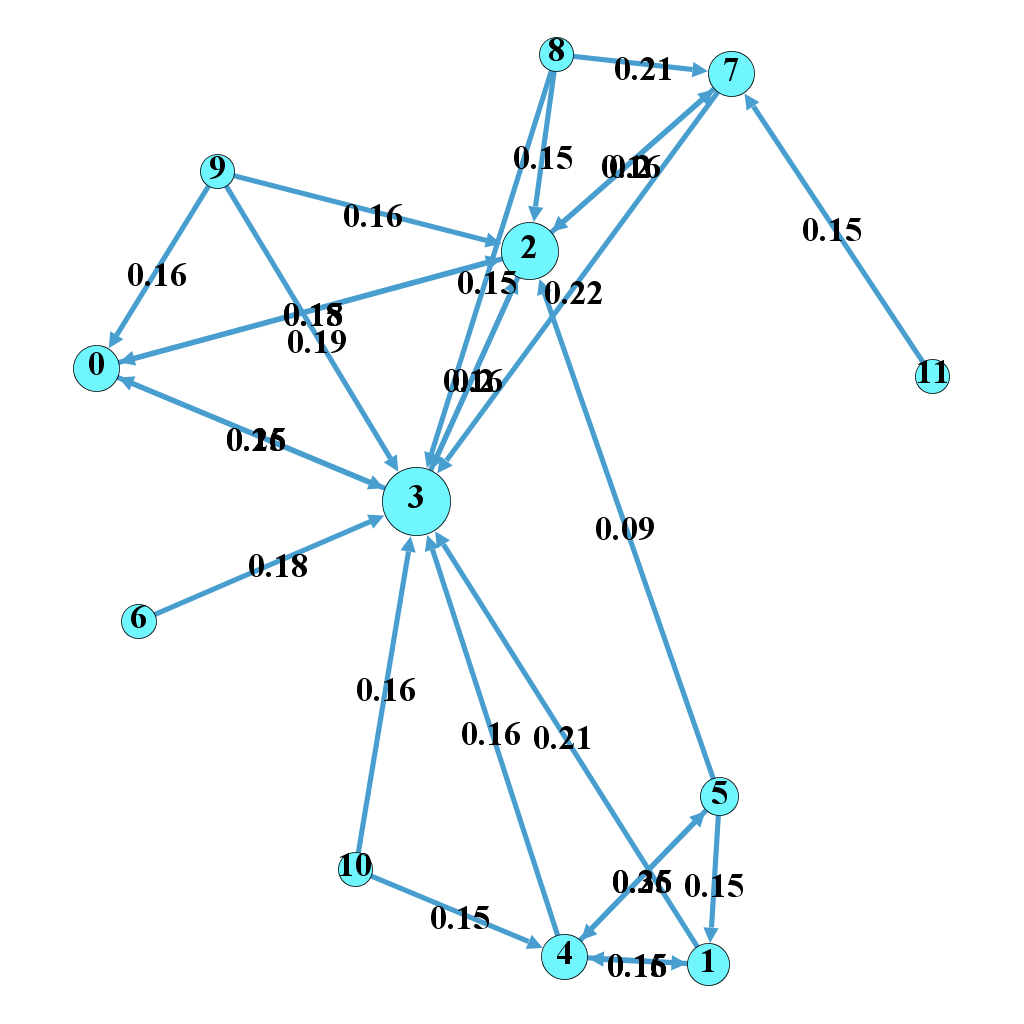}
\vspace{-0.70cm}
\caption{Discovered HMM consisting of $12$ states}
\label{fig:hmm}
\end{figure}
The user strategies are hidden and not directly observable (unlike activities). Figure \ref{fig:hmm} shows the topology of the discovered HMM and also some of the HMM parameters. We generated several HMMs models with different numbers of strategies but presenting the HMM with $12$ states only as a proof-of-concept due to limited space in the paper. As prescribed in the MMM, we eliminate the elements of the transition matrix $\mathbb{T}$ which are smaller than a specified threshold $\epsilon$=$0.15$. The HMM in Figure \ref{fig:hmm} consists of $12$ nodes where the size of the node is proportional to the number of incoming links. The labels in the edge (value $\geq$ $0.15$) represents the transition probabilities. The thickness of the edge is proportional to the value of the transition probability.  
\subsection{Discovering Intention-Oriented Process Model}
\begin{figure}[t]
\centering
\includegraphics[width=0.41\textwidth]{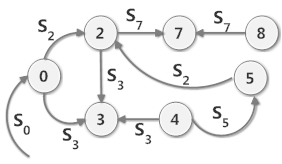}
\vspace{-0.66cm}
\caption{A section of the showing the assignment of strategies to sub-intentions}
\label{fig:map}
\end{figure}
Once the HMM and transition matrix $\mathbb{T}$ is derived, the next step consists of assigning each strategy in the matrix a target sub-intention. Each strategy leads to an intention according to the Map formalism. Next step of the method after discovering the HMM is to construct the relationship between strategies according to the transition matrix. Figure \ref{fig:map} shows a fragment of the Map generated as the result of connecting the strategies according to the procedure defined by the Map Miner Method. Strategy $\mathbb{S}_{2}$ is connected to strategy $\mathbb{S}_{3}$ and strategy $\mathbb{S}_{7}$ as there is an element in the transition matrix from $\mathbb{S}_{2}$ to $\mathbb{S}_{3}$ and $\mathbb{S}_{2}$ to $\mathbb{S}_{7}$. 

\begin{figure}[t]
\centering
\includegraphics[width=0.58\textwidth]{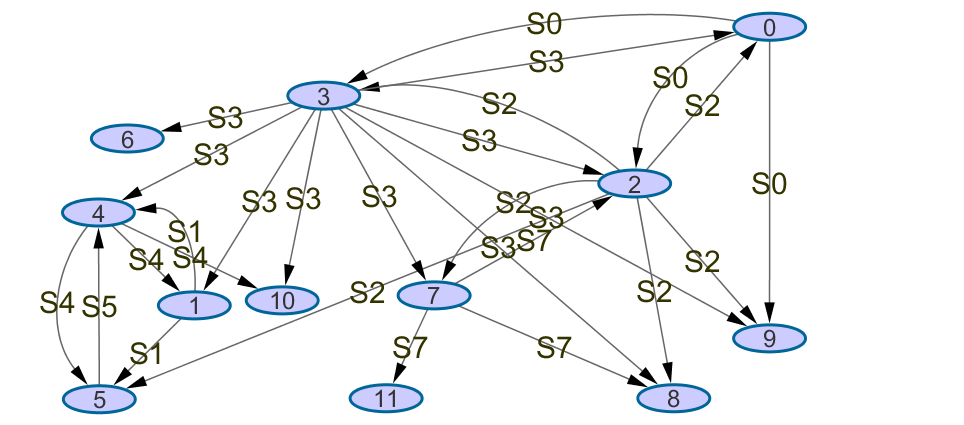}
\vspace{-0.53cm}
\caption{Complete Pseudo-Map consisting of strategies connected to sub-intentions}
\label{fig:pmap}
\end{figure}

\begin{table}[t]
\caption{Result of clustering sub-intentions to intentions. CC: Clustering Coefficient, CL: Closeness Centrality, EC: Eccentricity, NC: Neighborhood Connectivity}\label{tab:cluster} 
\begin{center}
\begin{tabular}{|l|l|l|l|l|l|}
\hline
\textbf{Node} & \textbf{Cluster} & \textbf{CC} & \textbf{CL} & \textbf{EC} & \textbf{NC} \\ \hline
11 & C1,C2 & 0.0 & 0.5 & 3 & 4.0 \\ \hline
6 & C1 & 0.0 & 0.5 & 3 & 9.0 \\ \hline
3 & C1, C2, C3 & 0.15 & 0.75 & 2 & 3.22 \\ \hline
2 & C1, C2, C3 & 0.23 & 1.0 & 1 & 4.16 \\ \hline
4 & C1, C2 & 0.25 & 0.6 & 3 & 4.25 \\ \hline
5 & C1, C2 & 0.33 & 0.66 & 2 & 4.33 \\ \hline
7 & C1, C2 & 0.33 & 0.75 & 2 & 4.75 \\ \hline
1 & C1, C2 & 0.5 & 0.54 & 3 & 5.33 \\ \hline
10 & C1 & 0.5 & 0.53 & 3 & 6.5 \\ \hline
0 & C1, C3 & 0.66 & 0.75 & 2 & 6.0 \\ \hline
8 & C1, C2 & 0.83 & 0.80 & 2 & 6.33 \\ \hline
9 & C1, C3 & 1.0 & 0.80 & 2 & 6.0 \\ \hline
\end{tabular}
\end{center} 
\end{table}

The next step consists of determining the Start and Stop intentions. The sub-intention(s) for which there is no incoming transition serves as the beginning of the process and corresponds to the Start intention. Similarly, the intention for which there is no outgoing transition corresponds to the end of the process and serves as the Stop intention. In our example, we have four nodes $6$, $8$, $9$ and $10$ for which there are no incoming transitions. We have node $11$ for which there is only one outgoing transition with transition probability of $0.15$. We experiment with node $6$ as the Start intention and node $11$ as the Stop intention. Figure \ref{fig:pmap} shows the complete pseudo-map consisting of all the strategies connected to sub-intentions. The pseudo-map shows how a sub-intention can be reached from other sub-intentions by following different paths or sequence of strategies. The model containing sub-intentions is called as pseudo-Map and are clustered in-order to group them into clusters of intentions. The parameter $N$ (number of clusters) determines the level of granularity and the number of intentions. After clustering a final Map process model is rebuilt. We use maximal clique-based EAGLE algorithm implemented in Cytoscape\footnote{ http://www.cytoscape.org/} to identify the clusters and group sub-intentions to intentions. Applying EAGLE algorithm is our context specific customization to the MMM. We use a clique-size threshold of $3$ and a complex-size threshold of $2$. Table \ref{tab:cluster} shows the result of clustering sub-intentions to intentions. We discover $3$ intentions and the Table \ref{tab:cluster} shows the nodes and strategies belonging to each of the intentions. 
\section{Threats to Validity}
The quality of discovered process map depends on the number of states defined in the HMM based on a heuristic and experience of an expert. The number of BWA iterations can also influence the outcome and in our case-study we set $50$ iterations which may not be optimal. Similarly, clustering sub-intentions to intentions requires parameters setting which may not be optimal for our case-study. 
\section{Conclusions}
Discovering strategies and intentions (intention-oriented process maps) allows understanding the service desk and IT operation behavior during the incident resolution process. The process model discovers multiple strategies and intentions and Table \ref{tab:case}, Figure \ref{fig:hmm}, \ref{fig:map} and \ref{fig:pmap} shows that service desk have selected different sequence of strategies or paths with different probabilities to fulfill their objectives or goals. We observe that activities External Vendor Assignment, Operator Update, Urgency Change, Communication with customer combined is a prevent strategy followed by several service-desk operators. Activties Reassignment, Communication with vendor, Assignment are part of one intention or cluster. A process owner can analyze the behavior of operators using the intention-oriented map in order to understand how, why and with which probabilities service desk operators address the reported incidents. The map also shows which paths are more or less taken are where are system bottlenecks. For example, network analysis reveals that the network diameter is $3$, network density is $0.318$, network centralization is $0.6$ and characteristic path length is $1.803$. Understanding best practices and best path to fulfill an intention is an important aspect of intention-oriented process mining. The discovered map connecting start state to end state and various sub-intentions and intentions can be used to provide recommendations to the operators in-terms of the best path to achieve a certain goal.

\bibliographystyle{abbrv}
\bibliography{comad2014}

\end{document}